# Simple absorbing boundary conditions for wave simulations with Smoothed Particle Hydrodynamics

Molteni Diego[1,*], Grammauta Rosario[1], Vitanza Enrico[2]
[1] Dipartimento di Fisica, Viale delle Scienze, Università di Palermo, Italy
[2] Dipartimento di Ingegneria Civile, Ambientale e Aerospaziale, Università di Palermo, Italy

**Abstract**
We study and implement a simple method, based on the Perfectly Matched Layer approach, to treat non reflecting boundary conditions with the Smoothed Particles Hydrodynamics numerical algorithm. The method is based on the concept of physical damping. We illustrate how it works in the case of 1D and 2D time dependent waves propagating in a finite domain.

## 1. Introduction

The problem of non reflecting boundary conditions is an old subject of the study of wave propagation in limited domains. The so called radiation boundary conditions at infinity have been studied since 1912 by Sommerfeld, but its practical implementation in computational solutions of electromagnetic field propagation can be referred to Engquist and Majda (1977). It is obvious that the occurrence of boundaries affects the evolution of a physical event that would otherwise propagate into open space. Many different strategies have bee adopted to circumvent the problem. Among numerous approaches the method of characteristics is well exploited in the fixed grid numerical methods (Poinsot and Lele, 1992). The perfectly matched layer (PML) approach, i.e. the use of an artificial absorbing layer, was devised by Berenger (1994) for simulations of electromagnetic waves and successively adopted in many wave field simulations: acoustics, seismic vibrations and fluids. For recent advancements for non linear regime of the Euler equations see Lin et al. (2011).

Recently Modave (2010) set up a simple and accurate method that is useful for linear and non linear shallow water simulations. The general idea is very simple. A sponge like absorbing layer is added to the physical domain. In this layer, sink or source terms are activated multiplied by a coefficient varying from zero, inside the physical domain, to a maximum at the outer edge of the sponge zone.
In general, the governing equations are written as follows:

$$\frac{dA}{dt} = f\left(A, \frac{\partial A}{\partial x}, x\right) - \sigma\left(A - A_{out}\right) \qquad (1.1)$$

[*] Corresponding author. Tel: +039091 6615055 Fax: +39091 6615069
E-mail address: diego.molteni@unipa.it

Where A is a generic fluid variable, $-\sigma\left(A - A_{out}\right)$ is the corresponding sink or source term, $A_{out}$ is the external boundary value, $\sigma$ is the damping coefficient different from zero only in the damping region. With an appropriate choice of the $\sigma$ spatial function this procedure produces extremely small reflection waves.
All these techniques are used for fixed grids discretization of the equations. In the lagrangean approach the characteristics approach has been suggested by Lastiwka et al. (2009). The PML approach is by far simpler, but, as far as we know, it has not be studied in the context of a lagrangean approach. We adopted this strategy for the lagrangean Smoothed Particle Hydrodynamics scheme and we tested it in the case of waves propagating in a finite pool. We show that the results are fairly good.
The Smoothed Hydrodynamics method (SPH) is a lagrangean mesh-free method based on a single basic interpolating function associated with each node of the moving mesh.
Here we give the basic ideas. For an up to date detailed presentation of the SPH method see Colagrossi e Landrini (2003). A function $f$ is interpolated from its known values at points $k$, by the approximation of the Dirac function integral:

$$f(x) = \int f(y)\delta(x - x')dy \Rightarrow \tilde{f}(x) = \sum_k f_k W(x, x'_k)\Delta x'_k \qquad (1.2)$$

Where $\tilde{f}(x)$ is the approximated function. Exploiting the mass density $\rho$, we can attribute to each moving node a mass $m_k = \rho_k \Delta x'_k$ and therefore the approximated function is given by:

$$\tilde{f}(x) = \sum_k m_k \frac{f_k}{\rho_k} W(x, x'_k) \qquad (1.3)$$

Consequently the space derivative can be approximated as:

$$\frac{\partial f}{\partial x} \approx \int \frac{\partial f(x')}{\partial x'} W(x - x')dx' \approx \sum_k \frac{m_k}{\rho_k} f_k \frac{\partial W_{ik}}{\partial x_i} \qquad (1.4)$$

Many details on the SPH approach can be also found in the review of Monaghan (2005). We give here the final formulae.
The continuity equation is given by:

$$\frac{d\rho}{dt} = -\rho\nabla\vec{v} \quad \Rightarrow \quad \frac{d\rho_i}{dt} = \sum_k m_k \left(\vec{v}_i - \vec{v}_k\right) \cdot \nabla_i W\left(\vec{r}_i, \vec{r}_k\right) \qquad (1.5)$$

The momentum equation:

$$\frac{d\vec{v}}{dt} = -\frac{1}{\rho}\nabla P \quad \Rightarrow \quad \frac{d\vec{v}_i}{dt} = \sum_k m_k \left(\frac{P_i + \Pi_i}{\rho_i^2} - \frac{P_k + \Pi_k}{\rho_k^2}\right)\nabla_i W\left(\vec{r}_i, \vec{r}_k\right) \qquad (1.6)$$

$P$ is the pressure to be given by an equation of state specific of the problem to be studied. It will be specified in the subsections. $\Pi$ is an artificial viscosity term needed to stabilize the equations (cfr. Monaghan 2005) and $W\left(\vec{r}_i, \vec{r}\right)$ is the interpolating function, named kernel, centered in the $\vec{r}_i$ point. This interpolating function has a scale factor $h$ and must have the properties to mimic the Dirac function, therefore:

$$\int_{-\infty}^{+\infty} W(x/h)\,dx = 1 \qquad \text{and} \qquad \lim_{h\to 0} W(x/h) = \delta(r)$$

The kernel used for our 2D simulation is the Wendland kernel function

$$W(r,h) = \frac{7}{4\pi h^2}\begin{cases} \left(1-\dfrac{r}{2h}\right)^4\left(1+2\dfrac{r}{h}\right) & if\ \dfrac{r}{h} \le 2 \\ 0 & if\ \dfrac{r}{h} > 0 \end{cases} \tag{1.7}$$

The integration in time is carried out by the predictor corrector algorithm for low Mach number flows, accurate to second order in the time step, proposed by Monaghan (2006). The time step is limited by the usual Courant condition.

## 2. The 1D shallow water case

The governing equations for shallow water waves, derived under the usual approximations of wave elevation much smaller than the full water depth, are well known. When written in the lagrangean form, and with constant water depth we have for the wave height:

$$\frac{dH}{dt} = -H\frac{\partial v}{\partial x} \tag{2.1}$$

Where $d.../dt$ is the comoving derivative, $H$ is the full height of the water level and $v$ is the fluid speed; the equation for the speed of the fluid is

$$\frac{dv}{dt} = -g\frac{\partial H}{\partial x} \tag{2.2}$$

Where $g$ is the gravitational acceleration.

These equations can be formally satisfied by an *fictitious* fluid having a density $\rho = H$ and an equation of state supplying the formula for the pressure $P = \frac{1}{2}g\rho^2$, so that the shallow water equation are fulfilled by this *special fluid* and therefore can be immediately approximated by the standard SPH formulae.

To damp appropriately the waves in proximity of the domain edge an extra spatial layer is added to the domain and the equations in this damping layer are:

$$\frac{dH}{dt} = -H\frac{\partial v}{\partial x} - \sigma(x)(H - H_0) \tag{2.3}$$

for the water level, and

$$\frac{dv}{dt} = -g\frac{\partial H}{\partial x} - \sigma(x)(v - v_0) \tag{2.4}$$

for the speed of the fluid,

where $\sigma(x)$ is the damping coefficient, which is function of the position in the layer, having an appropriate spatial dependence (discussed below); $v_0$ is the outflow speed. In our study we impose $v_0 = 0$ since in our case waves are propagating in a closed water pool. The $\rho_0$ is the reference density, it corresponds to $H_0$.

To produce a damping layer we add the following terms:

1. $S=-\sigma(x)(H-H_0)$ to the density equation

2. $Q=-\sigma(x)(v-v_0)$ to each component of the momentum equation.

We tested for the coefficient $\sigma$ the following functions suggested by Modave (2010) $\sigma=\sigma_0\left[\frac{(x-x_0)}{L}\right]^m$ where m is a positive integer, and $\sigma=\sigma_0\left[\frac{x-x_0}{(x_0+L)-x}\right]$ where $x_0$ is the starting point of the sponge zone. *L* is the amplitude of the damping layer and *m* is an exponent to be tuned.

Furthermore we tested also some *ad hoc* treatment, we call them *switches*, based on physical intuition:

1. Decrease the horizontal pressure force, only in the damping layer, with particular functions $f_1$ or $f_2$.

$$f_1(x) = -\frac{x-(x_0+L)}{L} \qquad f_2(x) = \frac{L^2-(x-x_0)^2}{L^2}$$

$f_1$ is a linear function, $f_2$ is a parabolic function with its maximum at $x_0$. We call them, *killing functions* since they reduce to zero the horizontal force acting on particles close to the end of the damping layer.

2. Use the damping friction: $\sigma>0$ only if $v_x<0$.

In a domain of amplitude $X=500$ we produce a Gaussian pulse in the density profile and a corresponding fluid speed according to the following prescription:

$$H(x)=H_0\left(1+0.01\exp\left(-\frac{(x-x_0)^2}{A^2}\right)\right) \qquad (2.5)$$

$$v(x)=(H(x)-H_0)\sqrt{gH_0} \qquad (2.6)$$

So we have a 1D soliton traveling towards the right side of the domain. The following pulse parameters have been chosen $H_0=1, x_0=3/4X$ and A=9h. The interpolating particle size is $h=2$, the particle spacing is $\Delta x=1$.

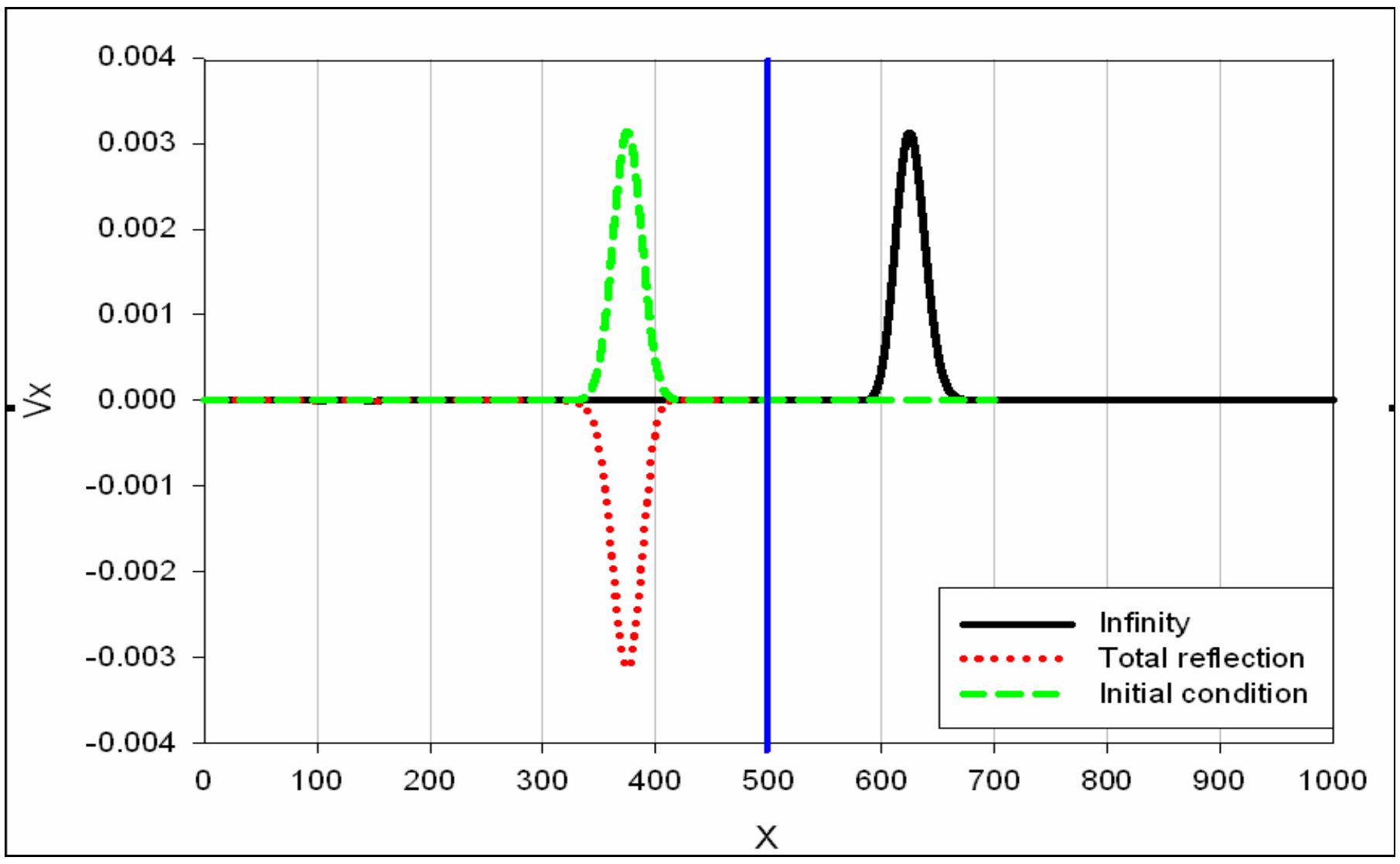

Fig. 1. Initial pulse profile chosen $H_0 = 1, x_0 = 3/4X$ and A=9h green dashed line; the perfectly reflected pulse: red dotted line and the infinity case: black line

Fig. 1 shows the initial analytical configuration and both the totally reflected Gaussian and the infinity propagated one. The reflection ratio for this set of 1D simulations is computed following Modave (2010) formula, i.e. the ratio of the errors

$$R = \sqrt{\frac{E_{lay,\infty}}{E_{refl,\infty}}} \tag{2.7}$$

given by:

$$E_{refl,\infty}(t) = \frac{1}{2} g \int \left(H_{refl} - H_\infty\right)^2 dx + \frac{1}{2} H \int \left(v_{refl} - v_\infty\right)^2 dx \tag{2.8}$$

Where the label $\infty$ identifies the values obtained with an extremely far right edge, i.e. no boundary condition (BC), the label *lay* refers to the quantities evaluated with a specific absorbing layer, the label *refl* refers to the quantities evaluated with a totally reflecting BC. Obviously the integrals have been replaced by a sum over the particles. Essentially we are measuring the differences of the flow variables and then we compute the relative energy, i.e. we are not computing the differences of the energies contained in the integration domain[1].

## *2.1. 1D Simulation Results*

Fig. 2 shows the reflection ratio for the various damping functions obtained with different values of the exponent *m* and the hyperbolic function. For small layers the best performances are obtained for the linear and the hyperbolic functions. However we focused our study on the hyperbolic function since it shows better results when we add the *ad hoc* physical switches.

---

[1] That would be $\frac{1}{2} g \int \left(H_l^2 - H_\infty^{\ 2}\right) dx + \frac{1}{2} H \int \left(v_l^2 - v_\infty^{\ 2}\right) dx$.

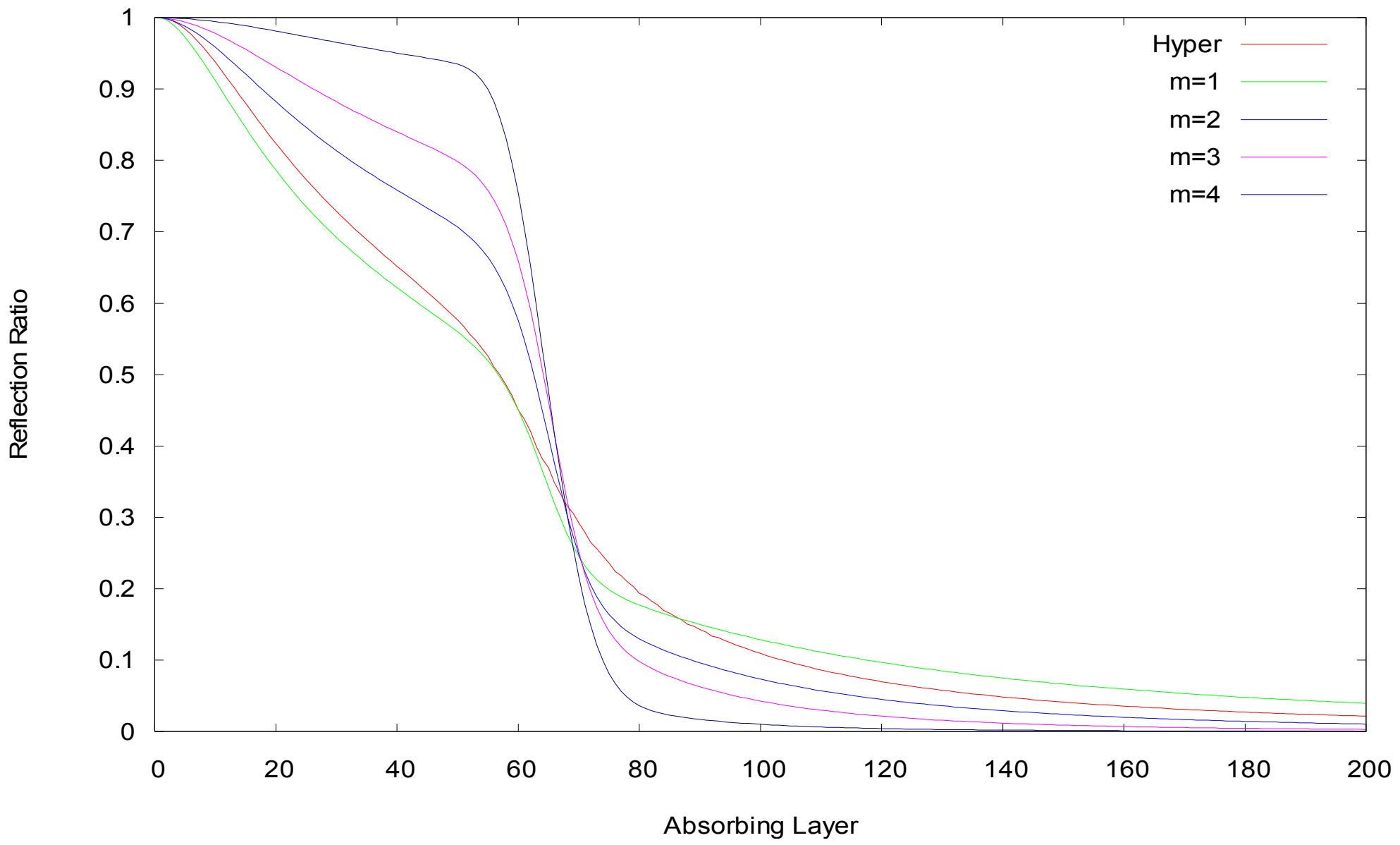

Fig. 2. Reflection ratio as a function of the thickness of the absorbing layer for different functions of the absorption coefficient. The parameters are identical to those used in Fig. 1.

Fig. 3 shows the reflection ratio as a function of the thickness of the absorbing layer for two different functions of the absorption (left panel: m=1, right panel: hyperbolic). The red line identifies the case with the use of the function without any switch, we call it the *pure* function; the green line identifies the results obtained with the same pure function with added a further tool: attenuation of the pressure force with a linear function $f_1$. The blue line identifies the case of attenuation of the pressure force with a quadratic function $f_2$. The best results are obtained with the green line, i.e. linear killing function.

Fig. 4 shows the reflection ratio results obtained adding the velocity switch. The best results are displayed with a blue line, corresponding to a damping with the use of the killing function $f_1$ and with the simultaneous use of unidirectional friction, i.e. use the damping friction $\sigma$ only if $v_x < 0$, so that the damping acts only if the speed of the particles (not of the wave) is negative.

Fig. 5 compares the best results obtained with the m=1 and the hyperbolic damping function using both the switches $f_1$ and $\sigma \neq 0$ *if* $v_x < 0$. The hyperbolic function works only moderately better, but in the 2D case we find a much better performance and therefore we focus on that function.

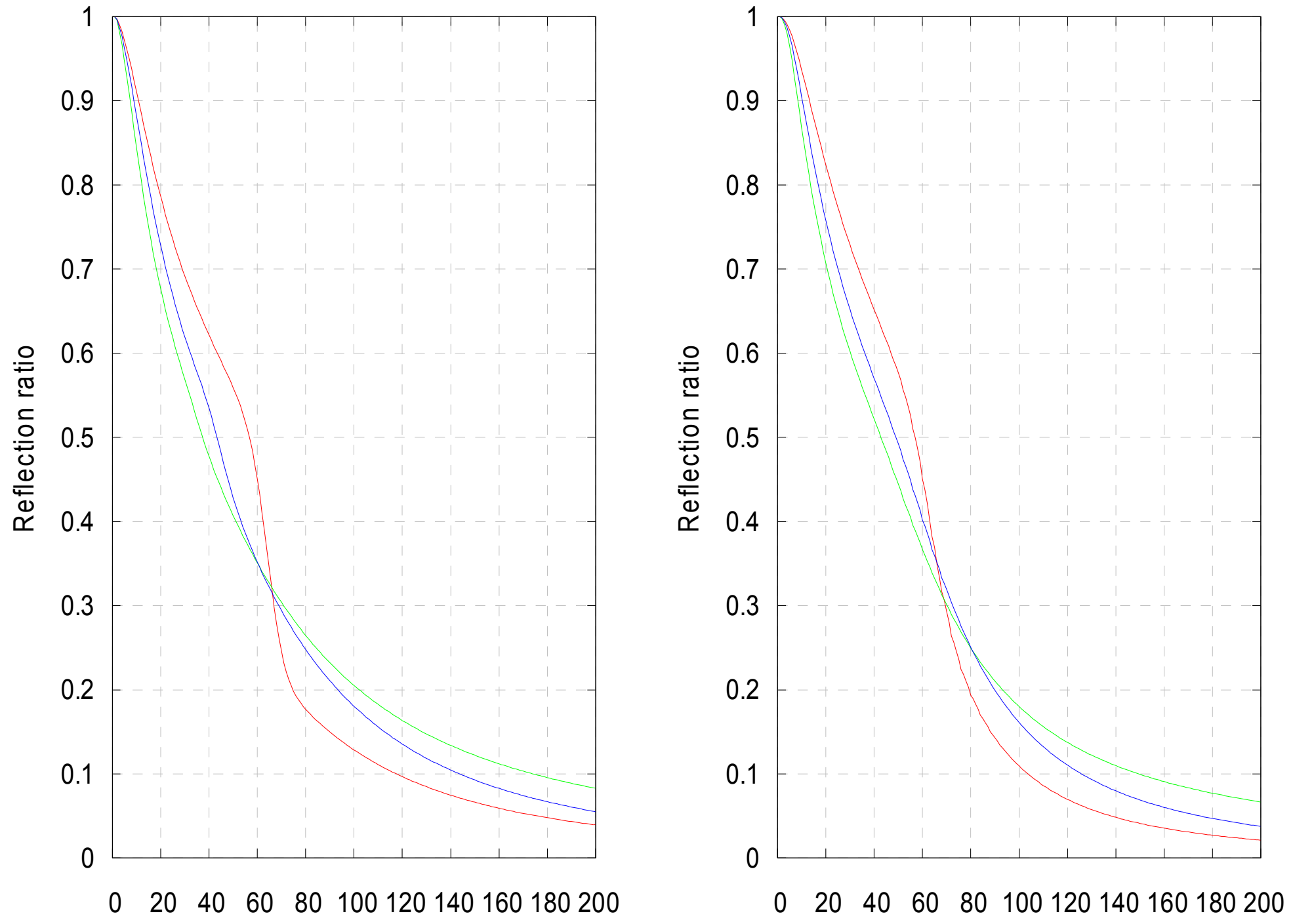

Fig. 3 Reflection ratio as a function of the thickness of the absorbing layer for two different absorbing coefficient functions (left panel: m=1, right panel: hyperbolic). The red line identifies the case with the use only the pure function, the green line identifies pure function multiplied by $f_1$. The blue line identifies the pure function multiplied by $f_2$. The parameters are identical to those used in Fig.1

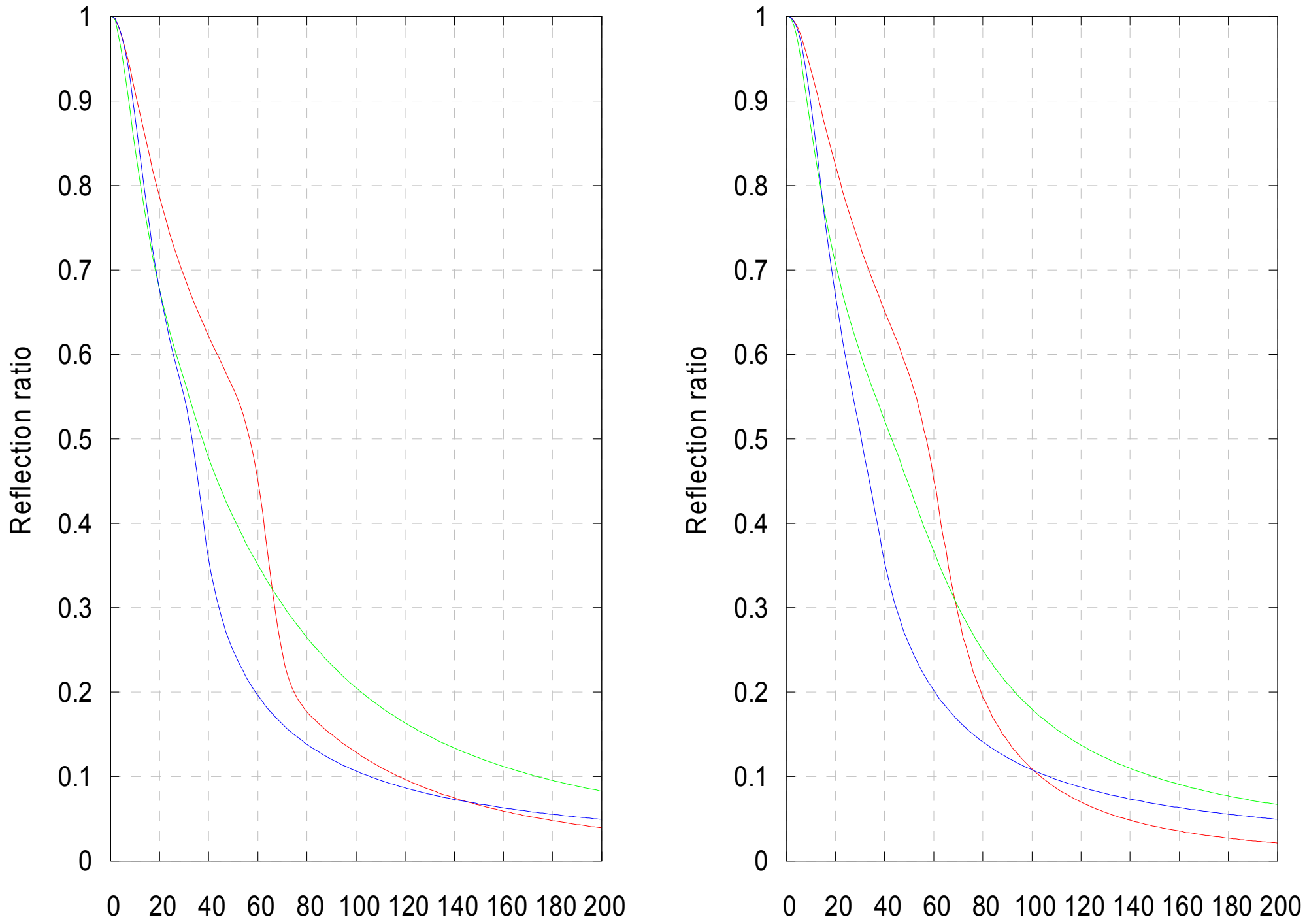

Fig. 4. Reflection ratio as a function of the thickness of the absorbing layer for two different distributions of the absorption coefficient (left panel: m=1, right panel: hyperbolic). The red line identifies the case with the use of the pure function, the green line identifies the same case with added the attenuation of the force with a linear function $f_l$ within the layer zone. The blue line identifies the previous case with a further switch on the speed. The parameters are identical to those used in Fig.1

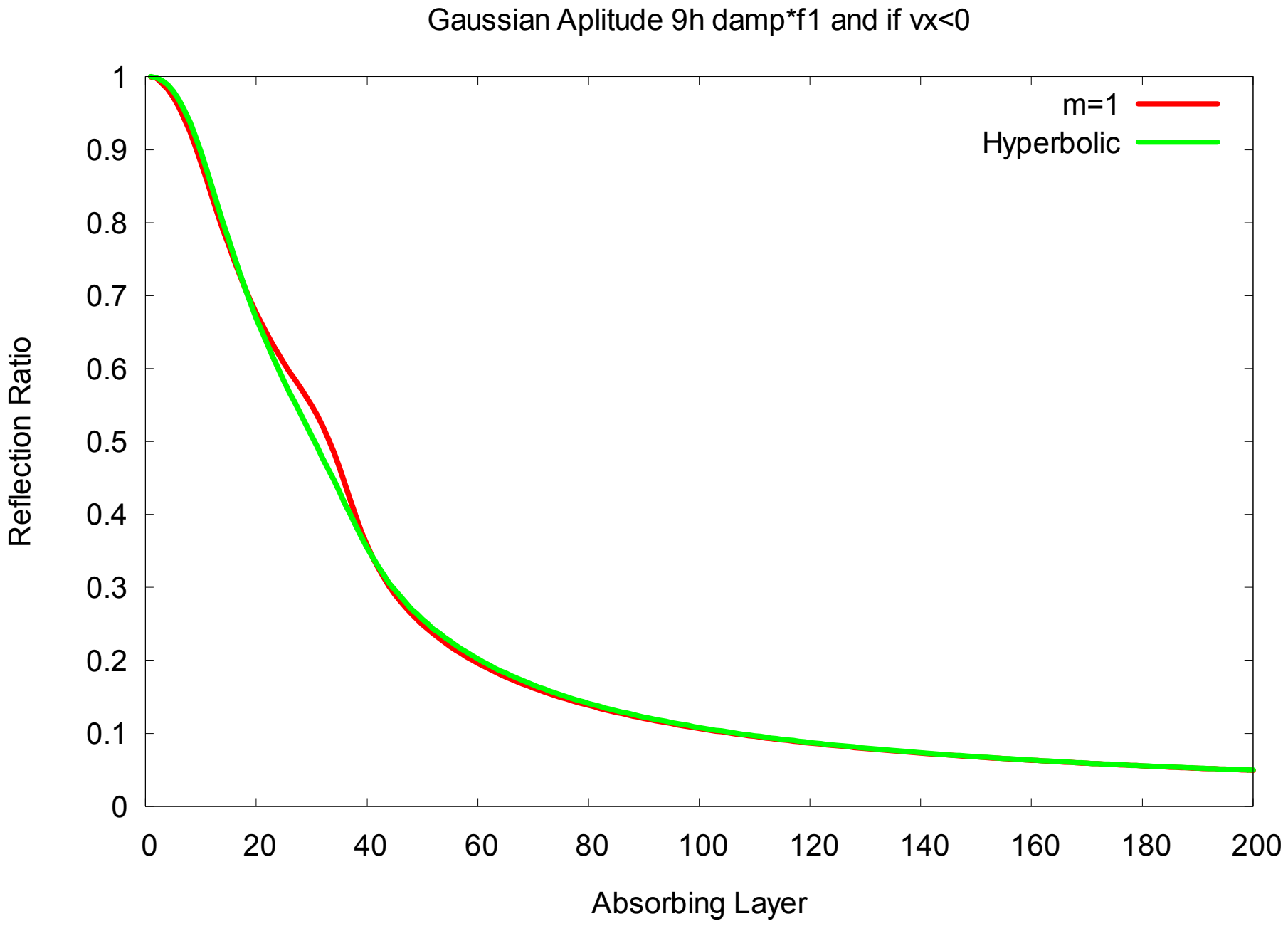

Fig. 5. Reflection ratio as a function of the thickness of the absorbing layer for two different distributions of the absorption coefficient (red line m=1 and green line hyperbolic). With attenuation of the horizontal pressure force with a linear function f1 within the damping zone and use damping only if the speed of the particles (not of the wave) is negative $v_x < 0$ in the absorbing layer at. The parameters are identical to those used in Fig.1

## *2.2. 1D Simulation Conclusion*

Resuming the 1D case, we may say that the best results have been obtained with the hyperbolic damping function $\sigma_0\left[\frac{x-x_0}{(x_0+L)-x}\right]$ plus two further treatments: the decrease of the horizontal pressure force with a linear function $f_l$ only in the damping layer and the use of damping only if the speed of the particles in the absorbing layer is negative $v_x < 0$.

We have to comment that, since we are using a lagrangean approach (the particles are free to move), we added in the denominator an extra softening term $0.5h$ to avoid division by zero if a particle reaches the left edge $\sigma = \sigma_0\left[\frac{x-x_0}{(x_0+L)-x+0.5h}\right]$. So hereafter we report only the results obtained with the hyperbolic damping function.

Finally Fig. 6 shows the values of the reflection coefficient versus the amplitude of the damping layer for different widths of the gaussian pulse. It shows the predictable result that the increase of the amplitude requires a larger damping layer to obtain the same reflection ratio.

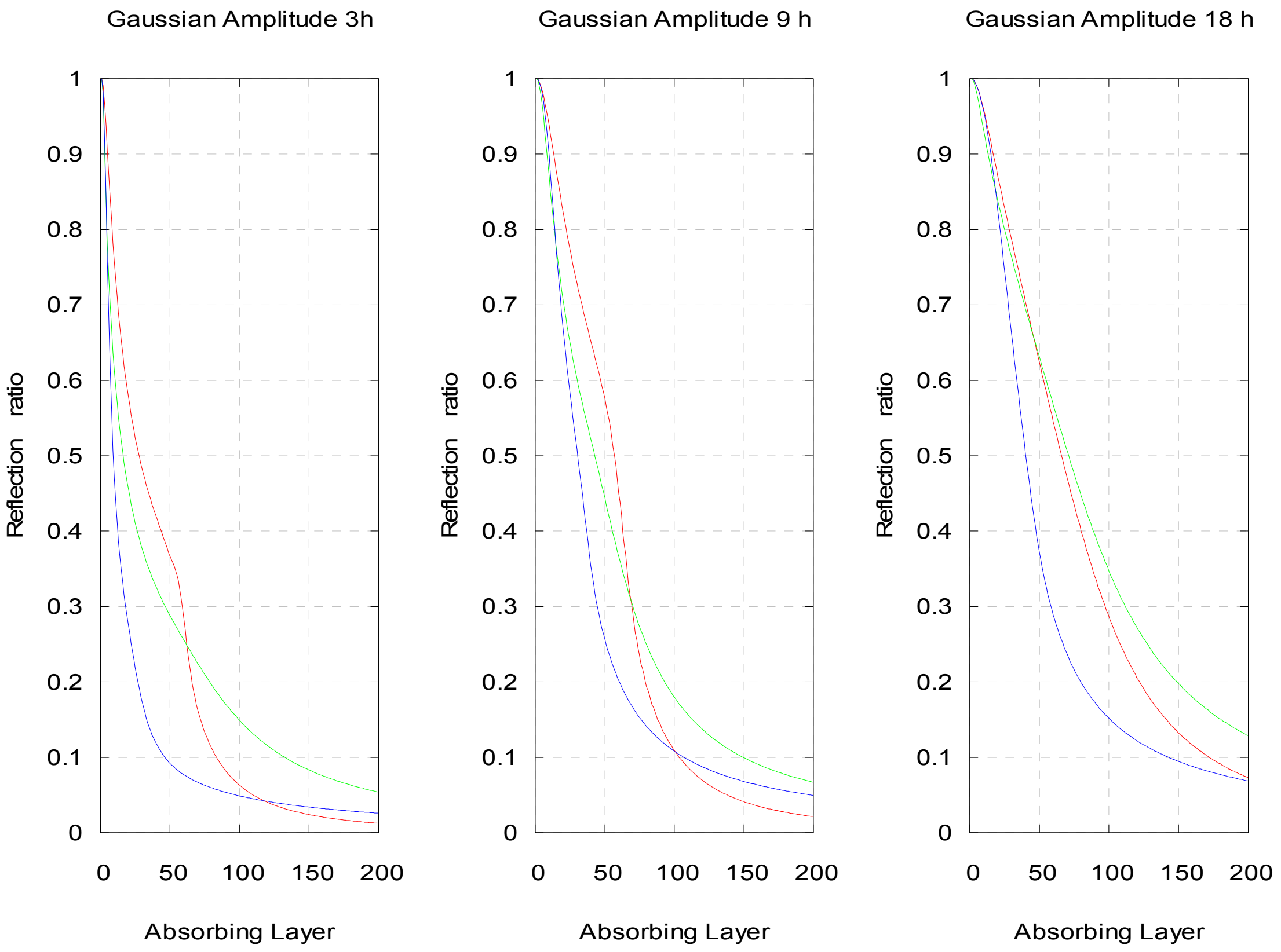

Fig. 6. Reflection coefficients for increasing length of the damping layer. The results displayed with the red, green and blue lines are obtained with the pure hyperbolic damping function, $f_1$ pressure factor , $f_1$ and speed switch respectively.

## 3. Waves in pool: 2D case

In this case we study the wavy motion produced by a wave maker palette in a water pool. The dynamics is truly two dimensional. To simulate incompressible water waves we use the weakly compressible approximation (Monaghan 2005), which consists essentially in the use of a sound speed an order of magnitude larger than the maximum typical speed of the water, we chose $c_s = 20v_{typ}$, where $v_{typ} = \sqrt{gH}$ . The governing equations are the previous ones, but with the equation of state given by the Tait equation:

$$P = \frac{\rho_0 c_s^2}{\gamma}\left[\left(\frac{\rho}{\rho_0}\right)^{\gamma} - 1\right] \tag{3.1}$$

with $\gamma = 7$.

The oscillating palette is placed at the left side of the rectangular pool. A damping layer is added in the right side. Check points of the water level are defined at regular space intervals. The setup is shown in Fig. 7.

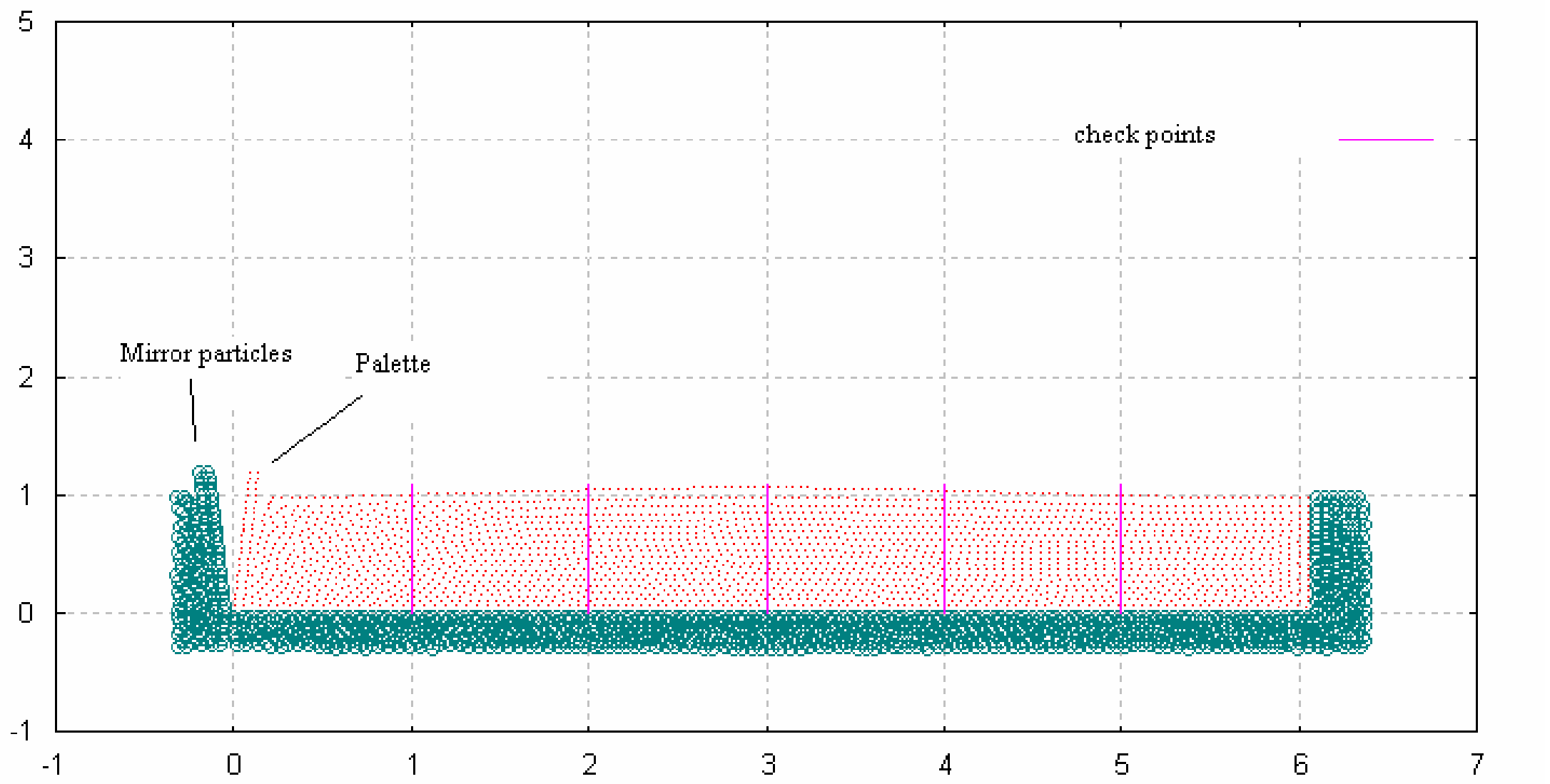

Fig. 7. Pool with palette and mirror particles, with no layer added.

The aim is to produce, in the finite pool of length X, a motion, unaffected by the right boundary, i.e. equal to the one obtained in the same zone but in an infinite pool.

We added the same source terms used for the 1D case to the equations of motion and continuity, taking into account the dimensionality of the problem, so we added a term $Q_\alpha = -\sigma_\alpha \left(v_\alpha - v_{0_\alpha}\right)$ to each component of the momentum equation. In this study $v_{0_\alpha} = 0$.

Also we explored some "*ad hoc*" terms, guided by the 1D experience and physical intuition:

- a switch on the damping triggered by the speeds $v_x$ and /or $v_y$.

-the use of killing functions to reduce the horizontal component of the forces due to the pressure.

We examined the case of a continuous periodic wave and that of a wave generated by a single sinusoidal oscillation.

*3.1. Continuous periodic wave*

We made a pool of length X=6.061 meters and height 1 m. The particles have an intrinsic width h=0.1 and are placed at a regular spacing $\Delta l = 0.05$ in X and Z. The number of particles in the pool is N=2570 (palette included). The boundaries are made with mirror particles procedure. The palette oscillates with a period P= 2.236969878 sec and with an angle amplitude of 5 degrees. With these values the water in the pools enters in a resonant state. Fig. 8 reports the levels of the water column, measured at five different positions (at x= 1,2,3,4,5 meters) along the pool, versus elapsed time, in the case of a very long "infinite" pool .

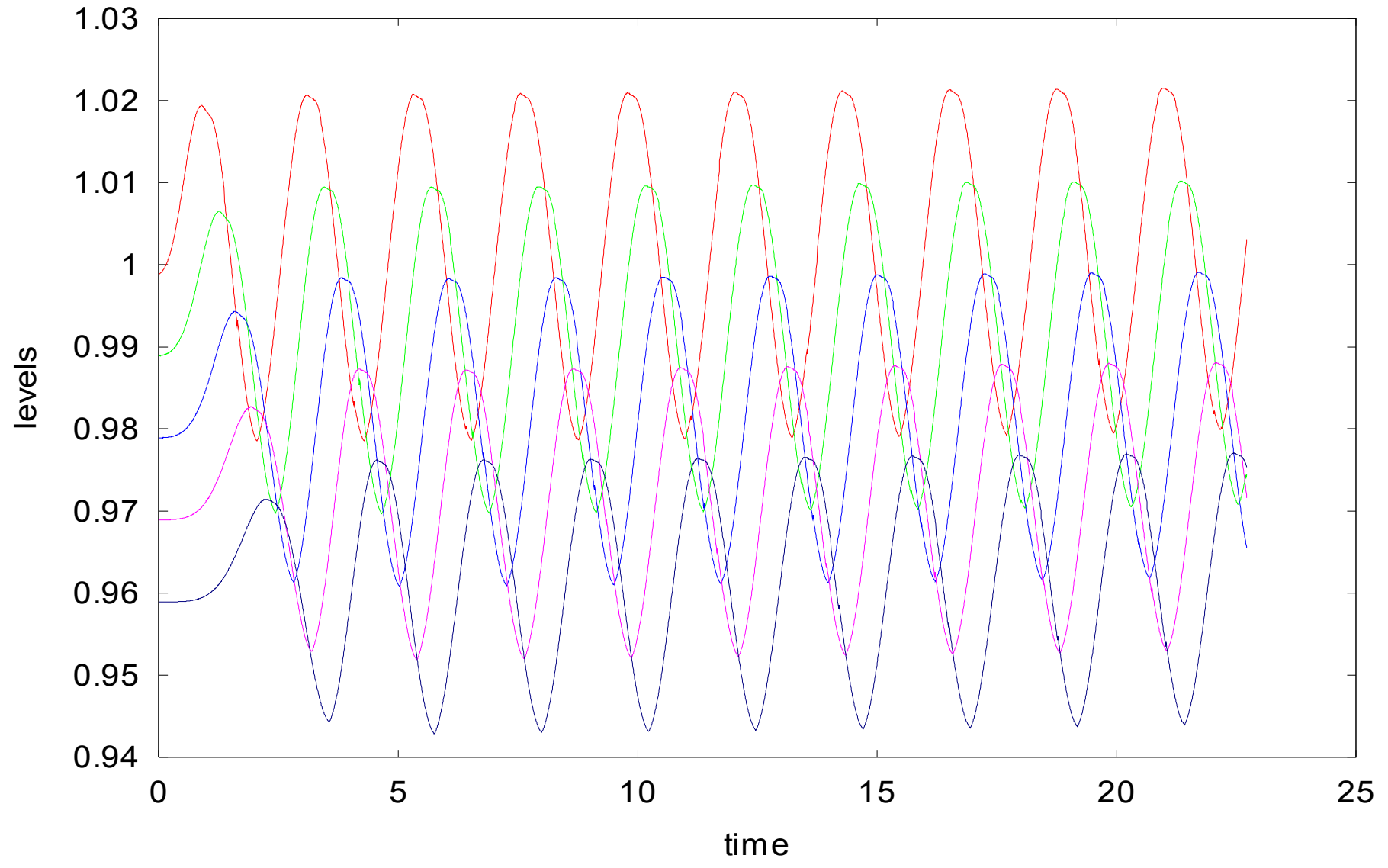

Fig. 8. Levels at x=1, 2, 3, 4, 5 m. in an "infinite pool"

The levels are vertically shifted for clarity. It is clear that the waves propagates without disturbances.

Fig. 9 shows the velocity field of the pool in resonant condition, with the palette and the mirror points.

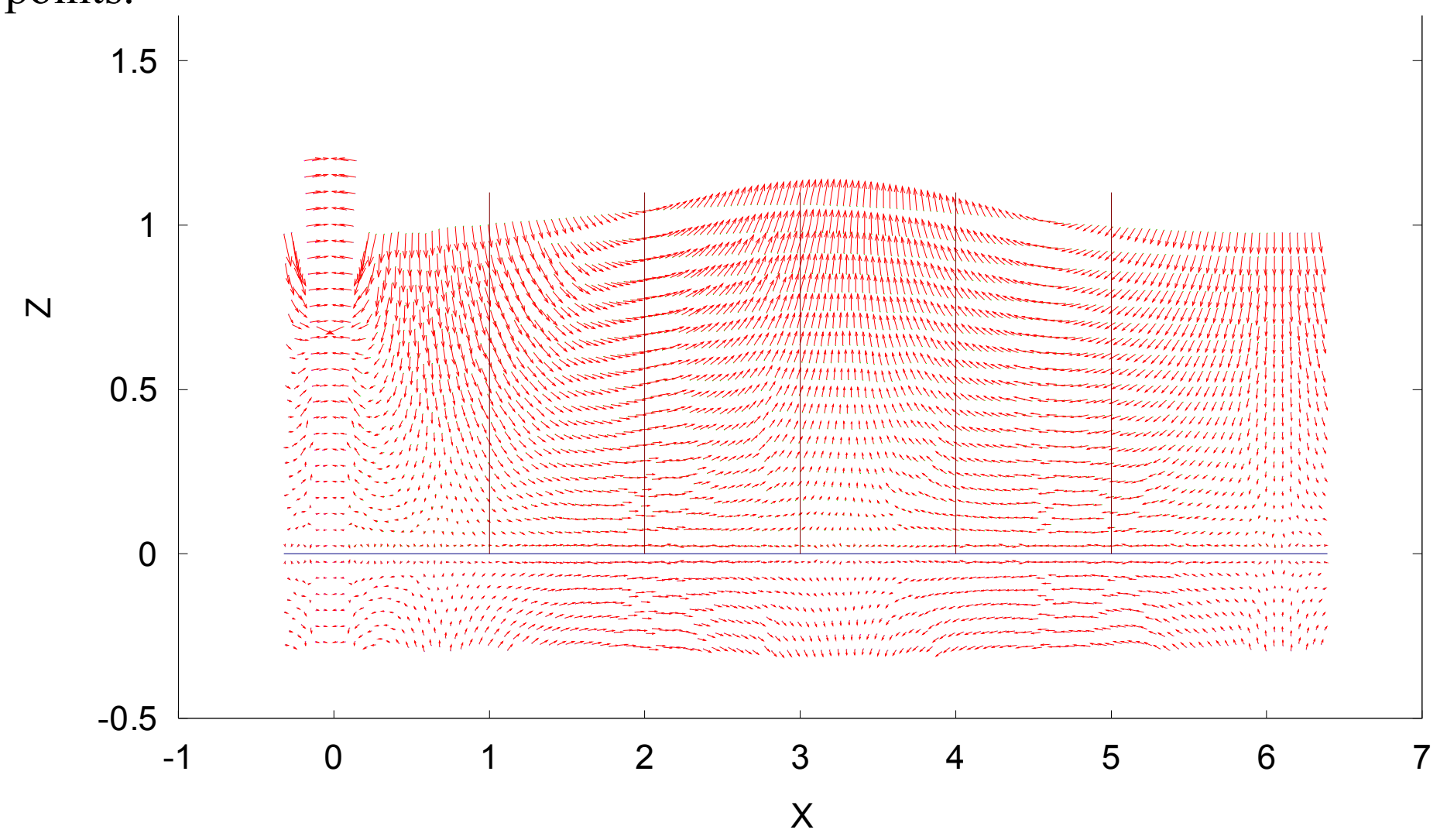

Fig. 9. Velocity field of the water in the resonant case, no damping layer

Fig. 10 shows the levels for the resonant pool. The levels are shifted in the Z coordinate by a small amount dz = 0.01 for clarity.

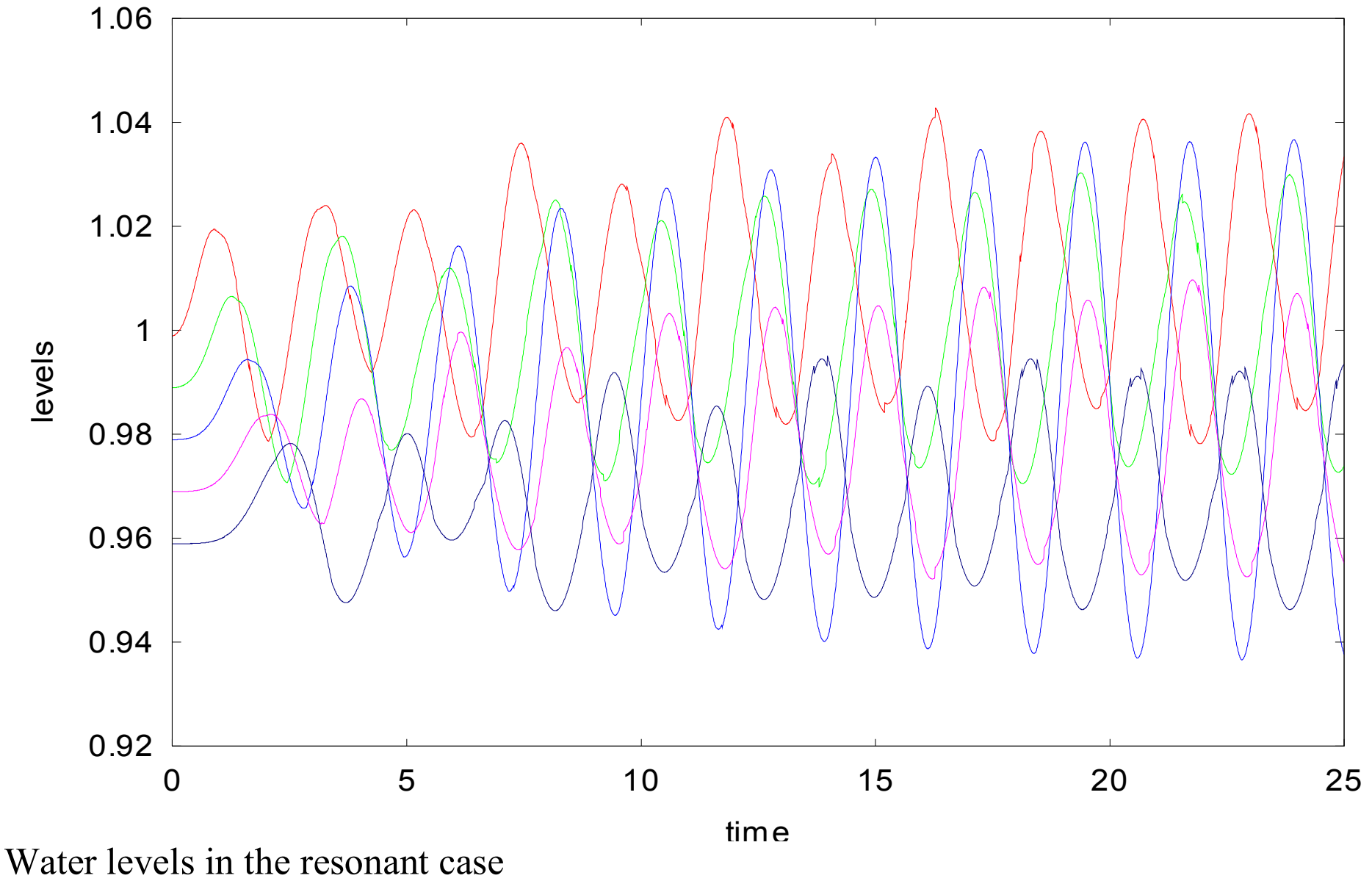

Fig. 10. Water levels in the resonant case

It is clear that the oscillations are larger and increasing with time.

In Fig. 11 we show the resulting levels when a damping layer of extension L=6.061 m is added. That is, we are using a full simulation domain X=12.122m. In all these tests the hyperbolic function $\sigma = \sigma_0 \left[ \frac{x - x_0}{(x_0 + L) - x + 0.5h} \right]$ is adopted. The coefficient $\sigma_0$ has the dimension of 1/time and we chose its value as $\sigma_0 = v_{ref} / L$. For this study the reference speed has been chosen equal to the sound speed: $v_{ref} = c_{sound}$ (case with the label *M0*).

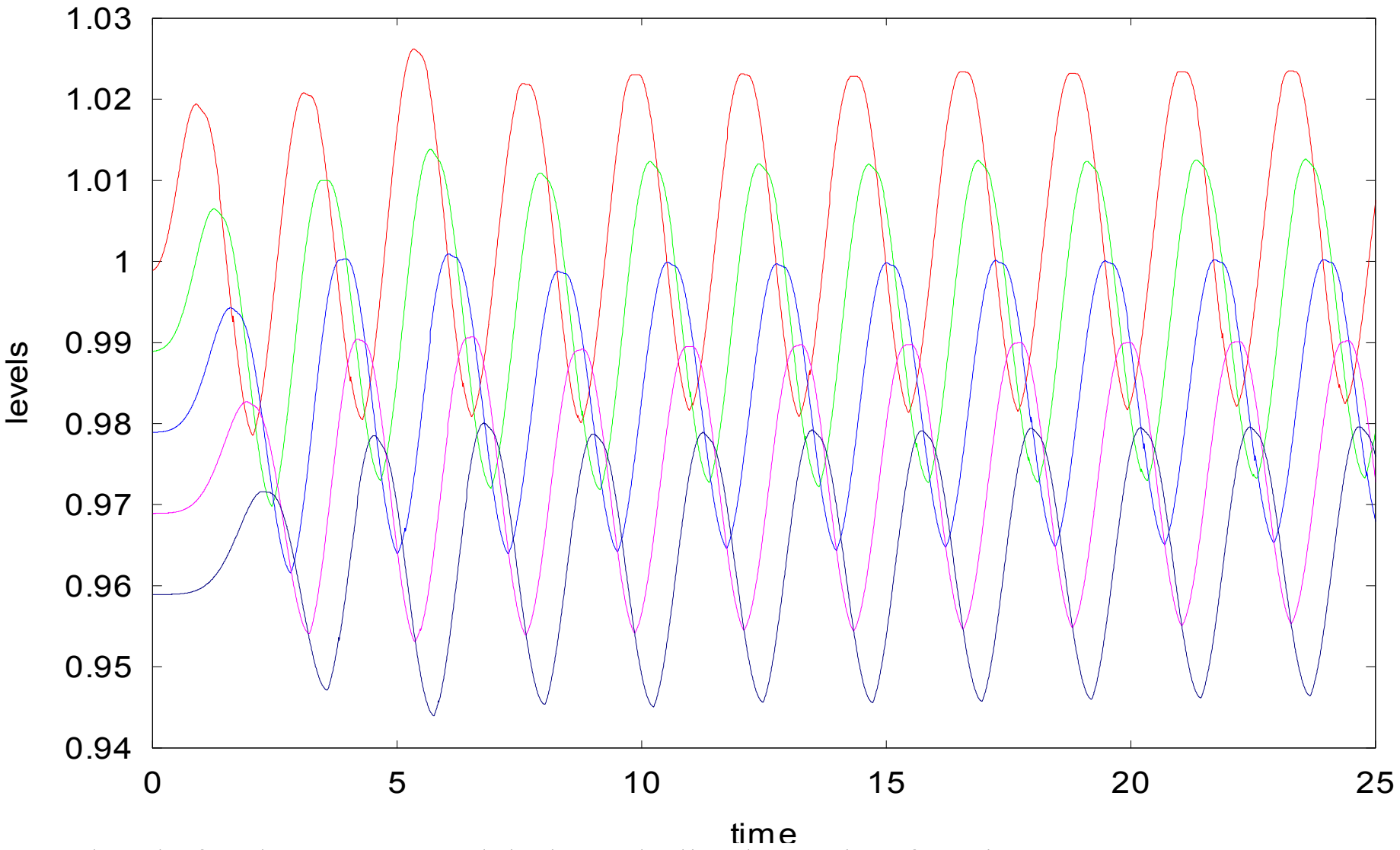

Fig. 11. Water levels for the case M0: plain hyperbolic absorption function.

It is clear that the wave profiles are very similar to the ones of the *infinity* case.

The kinetic energy content can be used as an indicator of the similarity of the flows. The Fig. 12 shows the kinetic energy of the water (computed excluding the damping layer contribution) versus time for various cases. It is clearly shown the increase of the energy in the resonant condition, the steady oscillating energy for the *infinity* case, together with the very close values obtained with different damping layer cases.

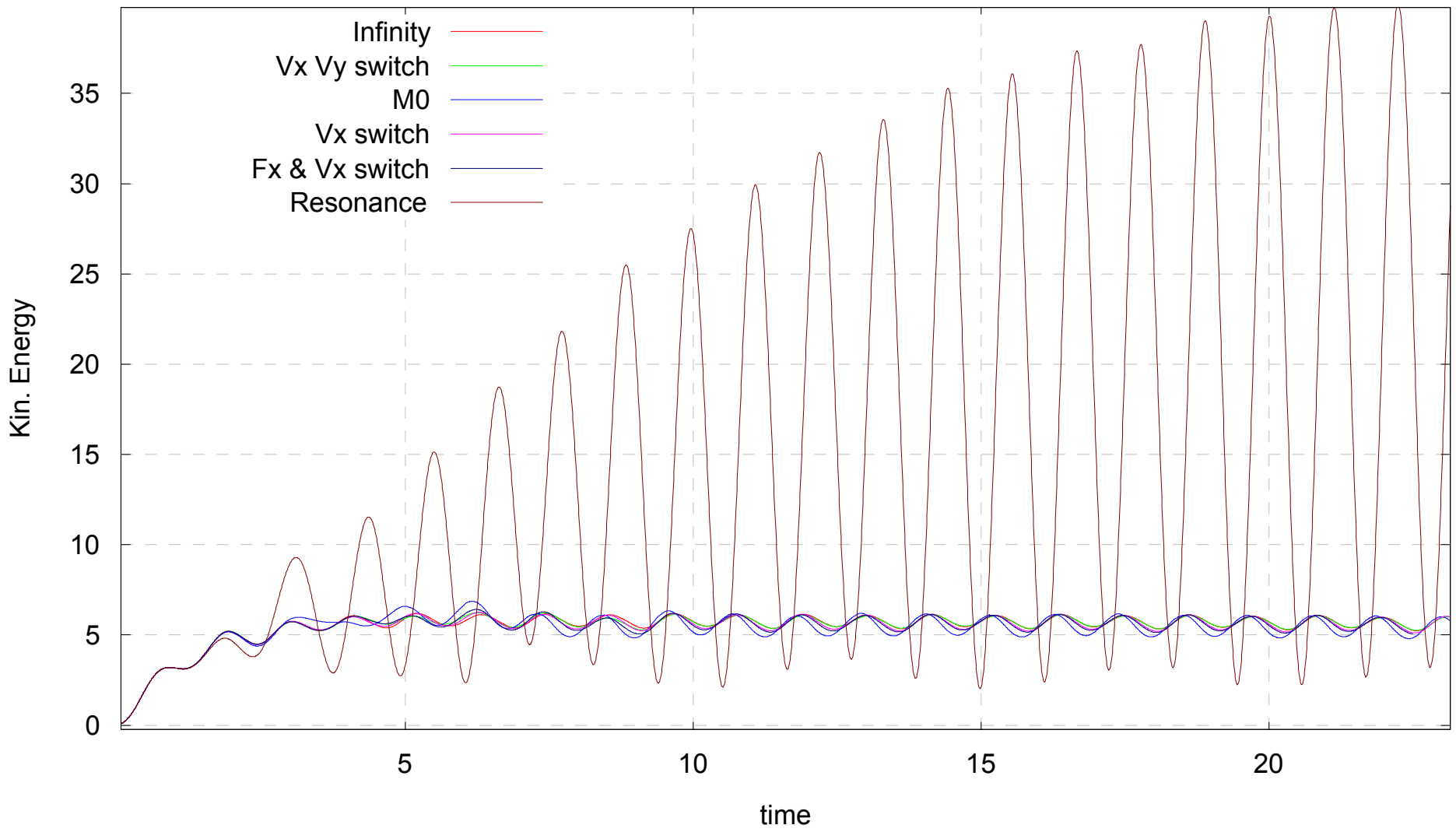

Fig. 12. Kinetic energy versus time for resonance and for boundary layer with switches

If we make a zoom of Fig. 12, on the values of the kinetic energy, we obtain the Fig. 13 and we find some discrepancies between the *infinite* case and the *M0* one.
We made the same simulation to test the following set of switches:
"$V_x$ $V_y$": the damping function works only if $v_x$<0 and $v_y$<0
"$V_x$": the damping function works only if $v_x$<0
"$F_x$ & $V_x$": the damping function works only if $f_x$<0 and $v_x$<0
If we look to the kinetic energy we find that performances better than *M0* are obtained with each of the switch options mentioned, since for *M0* the kinetic energy has oscillations larger than the ones of the *infinity* case. Qualitatively, the best result seems to be obtained with the simple "$V_x$" switch. However further investigations should be carried out to produce a numerical estimate.

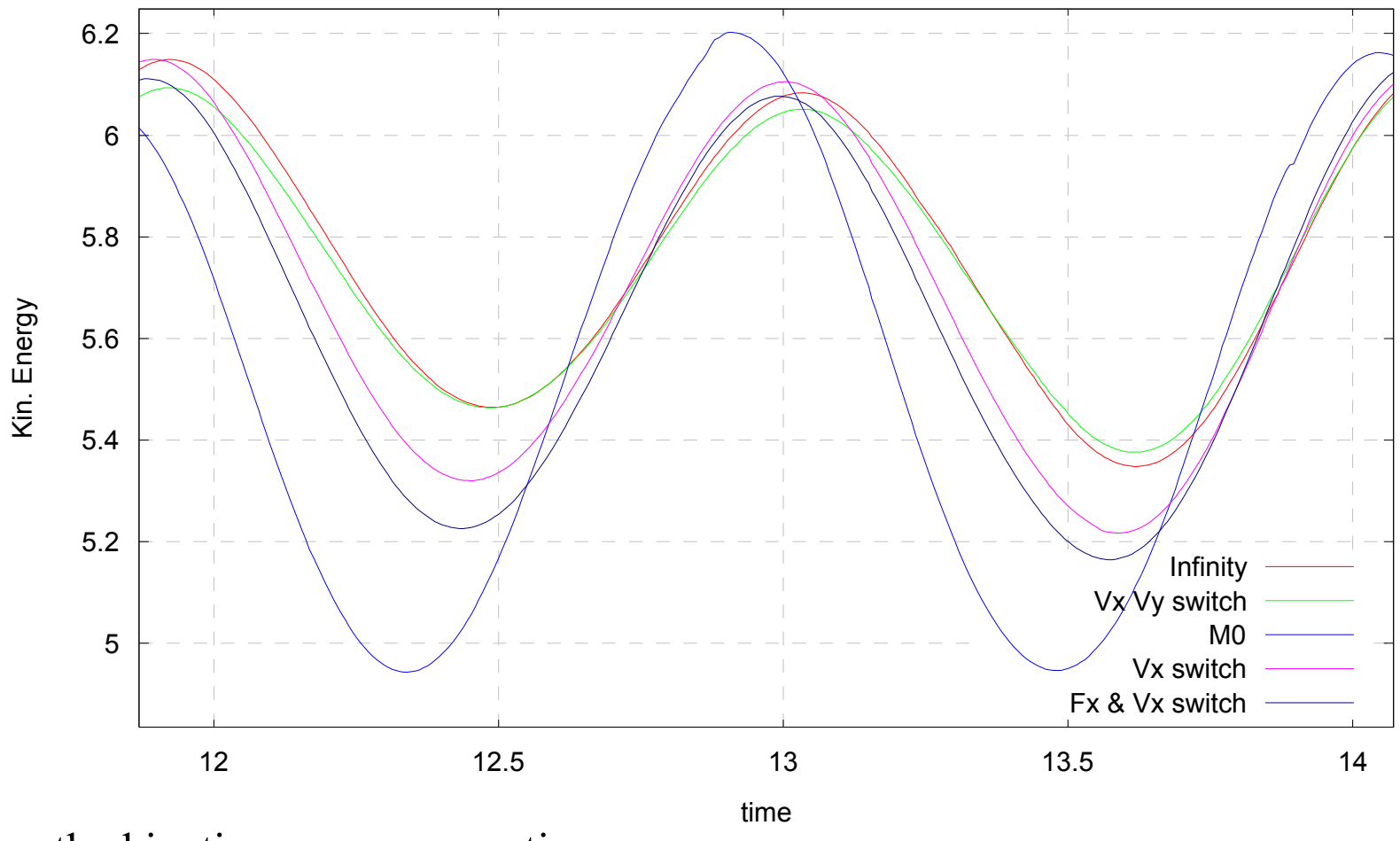

Fig. 13. Zoom on the kinetic energy versus time

*3.2. Sinusoidal single impulse wave*

We made also simulation of a single impulsive sinusoidal wave. In the same pool, the palette makes a single oscillation with the same angular amplitude and period of the previous simulation. In this case it is easy to see the effects of the reflected wave. To have a detailed information, in this case we chose ten elevation level points located at intervals of 0.6m, starting from 0.3m from the left side. Fig. 14 shows the particles and their velocity field for the water in the pool at time =25 sec; this is the configuration without any damping layer, i.e. pure reflection conditions at the right side. The vertical lines identify the points of level measurement. The particles distribution and the speed arrows show that the water level is still oscillating.

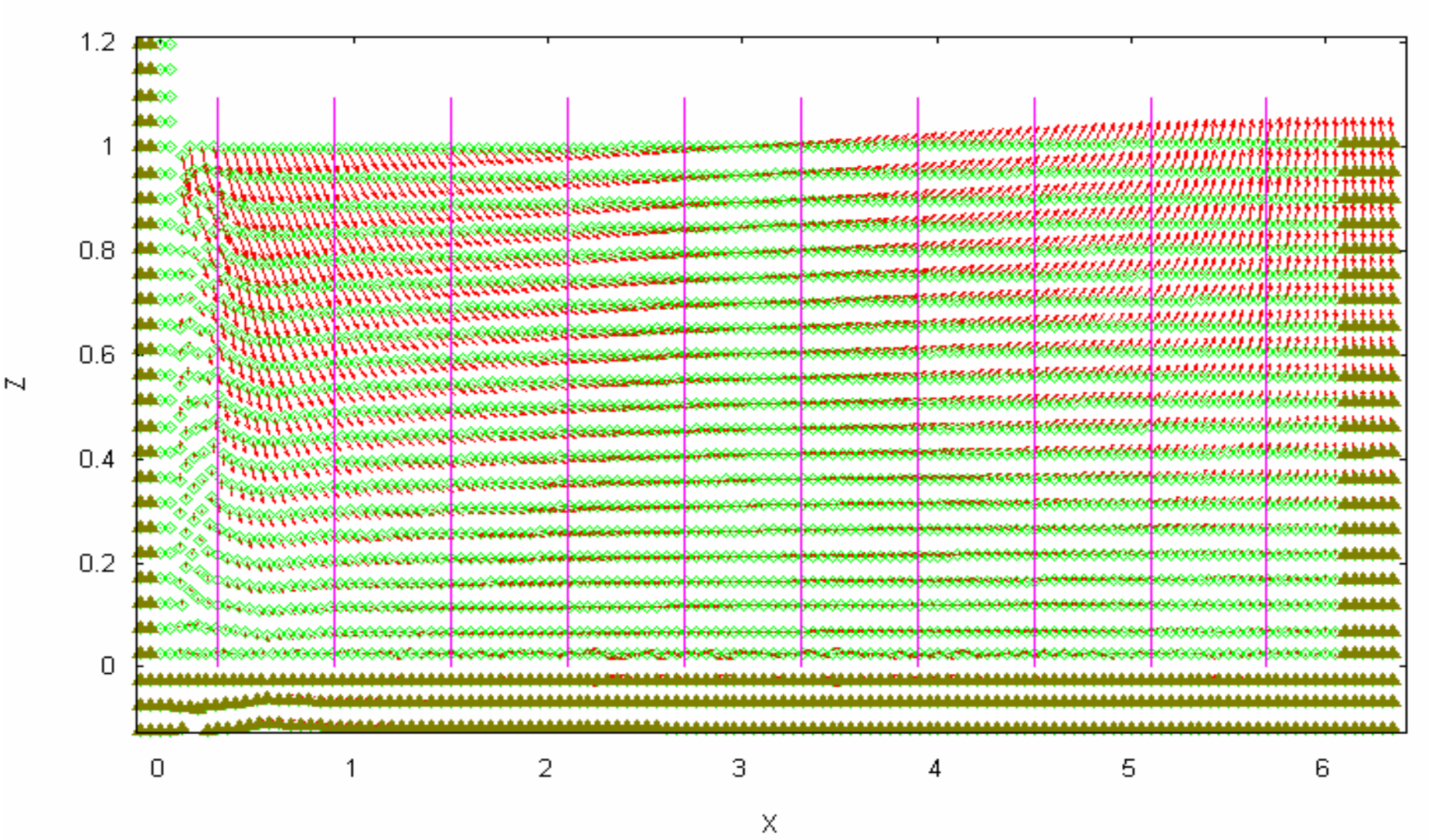

Fig. 14. The particles configuration and their speed (enlarged by a factor 2.5) for the water in a reflecting pool

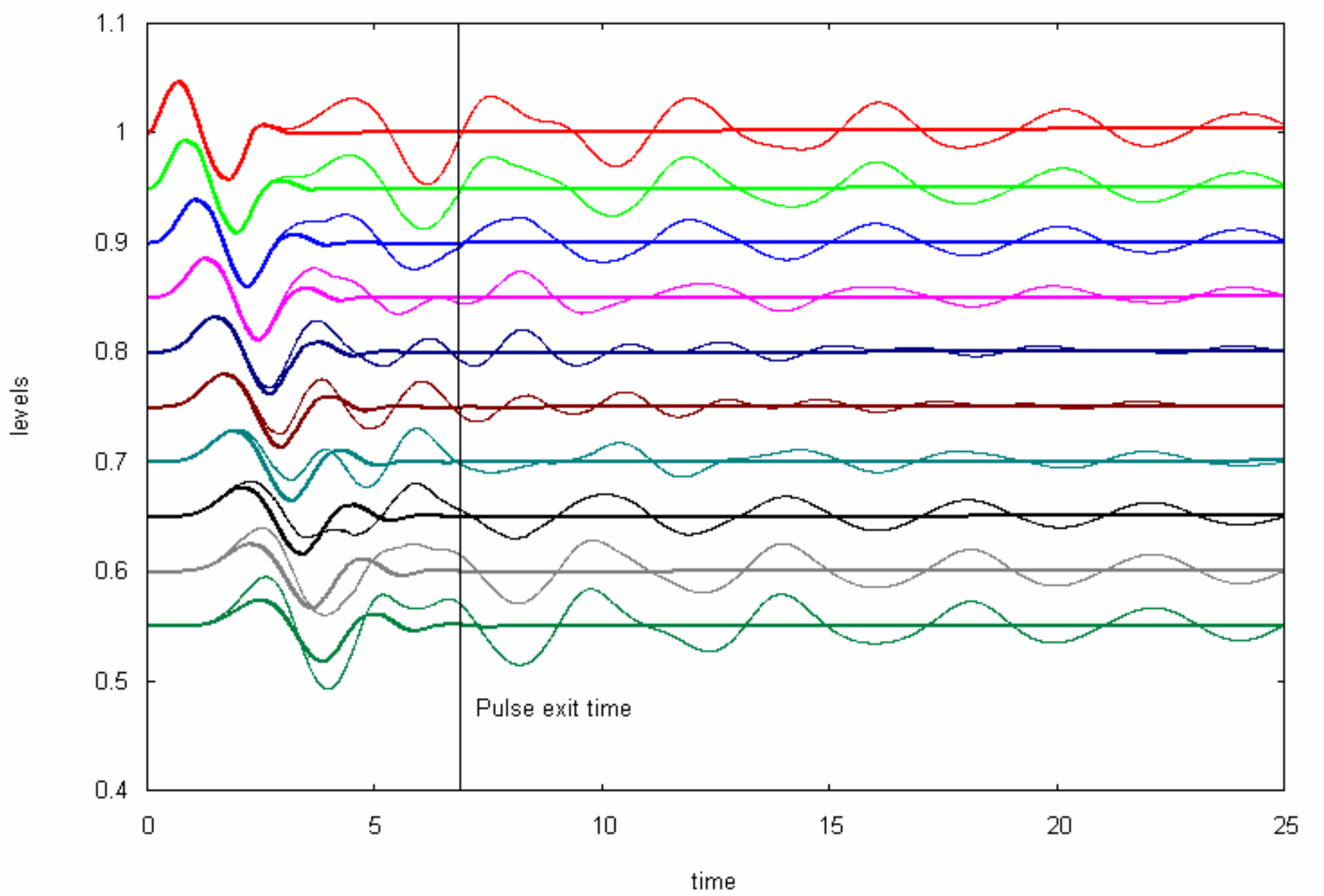

Fig. 15. Ten levels for reflecting BC pool compared to the levels of the *infinity* case

Fig. 15 shows in the same panel the levels when the pool has a simple reflecting boundary on the right side; together are plotted the levels obtained for the *infinite* pool (the thicker and straighter horizontal lines). It is obviously clear that oscillations are present even after long time the pulse had to be outside the pool.
In Fig. 16 are shown the levels obtained with the *M0* prescription compared with the *infinity* case.

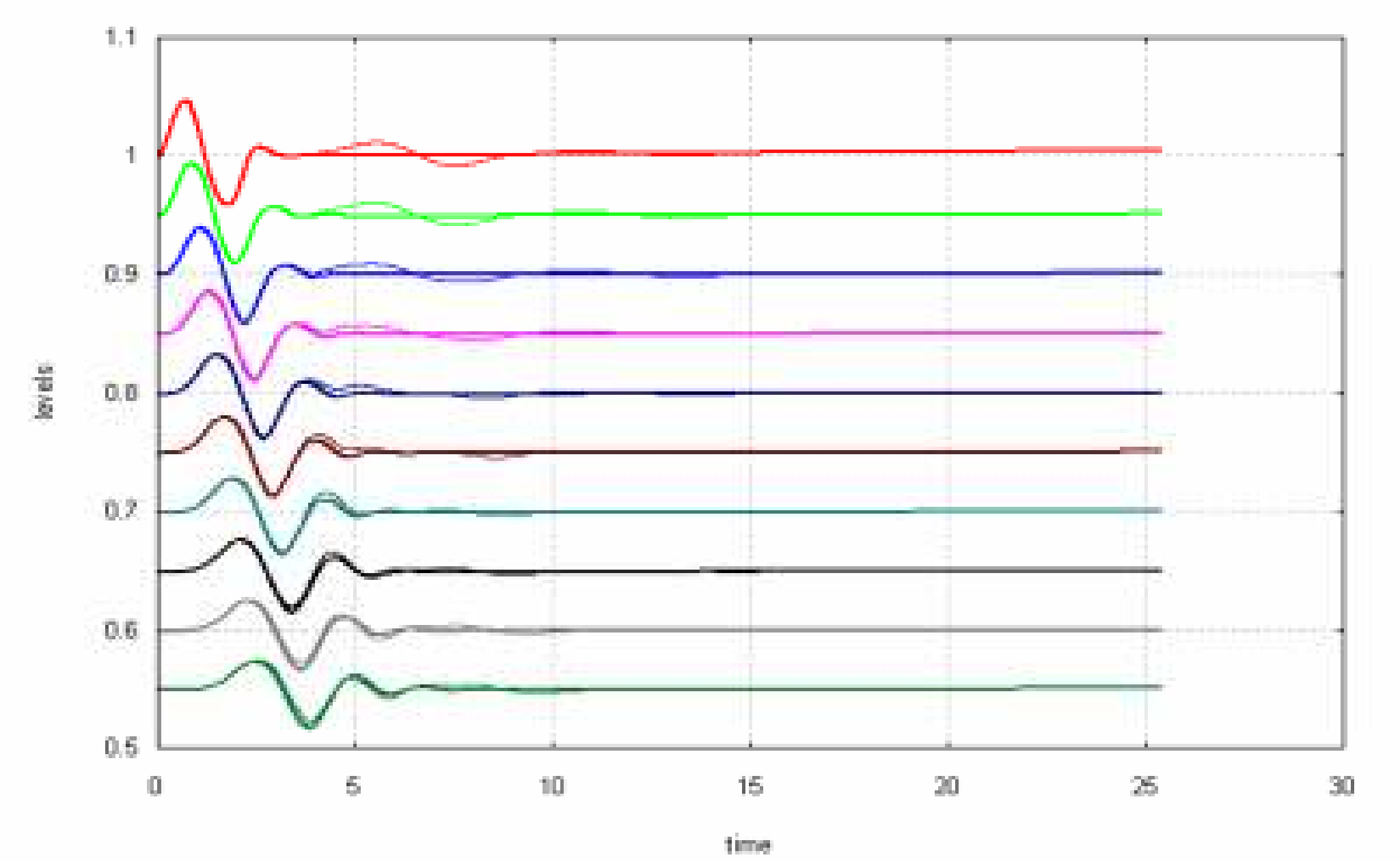

Fig. 16. Comparison of levels obtained with the *M0* prescription and the *infinity* case

In Fig. 17 the *M0* levels are plotted together with the ones obtained with the "Vx Vy" switch. The *M0* lines show a small bump around time $t = 5 \leftrightarrow 6$ sec, while the thick lines are more straight, they correspond to the use of the "Vx Vy" switch.

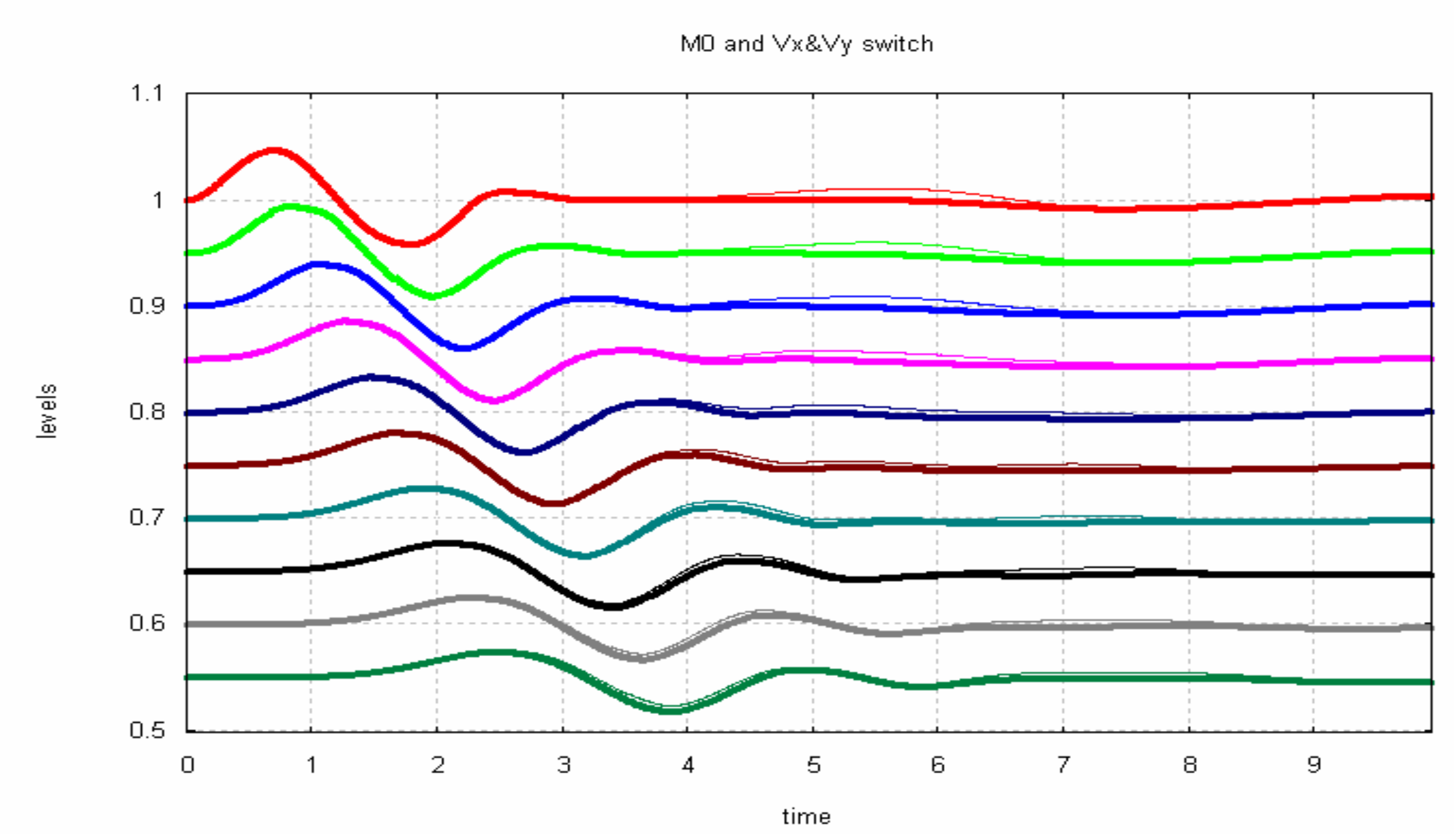

Fig. 17. Comparison of levels obtained with M0 and with the switch Vx&Vy (thick lines)

We studied also the same two problems using smaller damping layers. The results are similar to the one presented here with the obvious difference that the damping effects diminish as the sponge layer decreases. We show the results of the single sinusoidal impulse. Fig. 18 shows the kinetic energy of the pool versus time when the length of the damping zone is L=6.06.

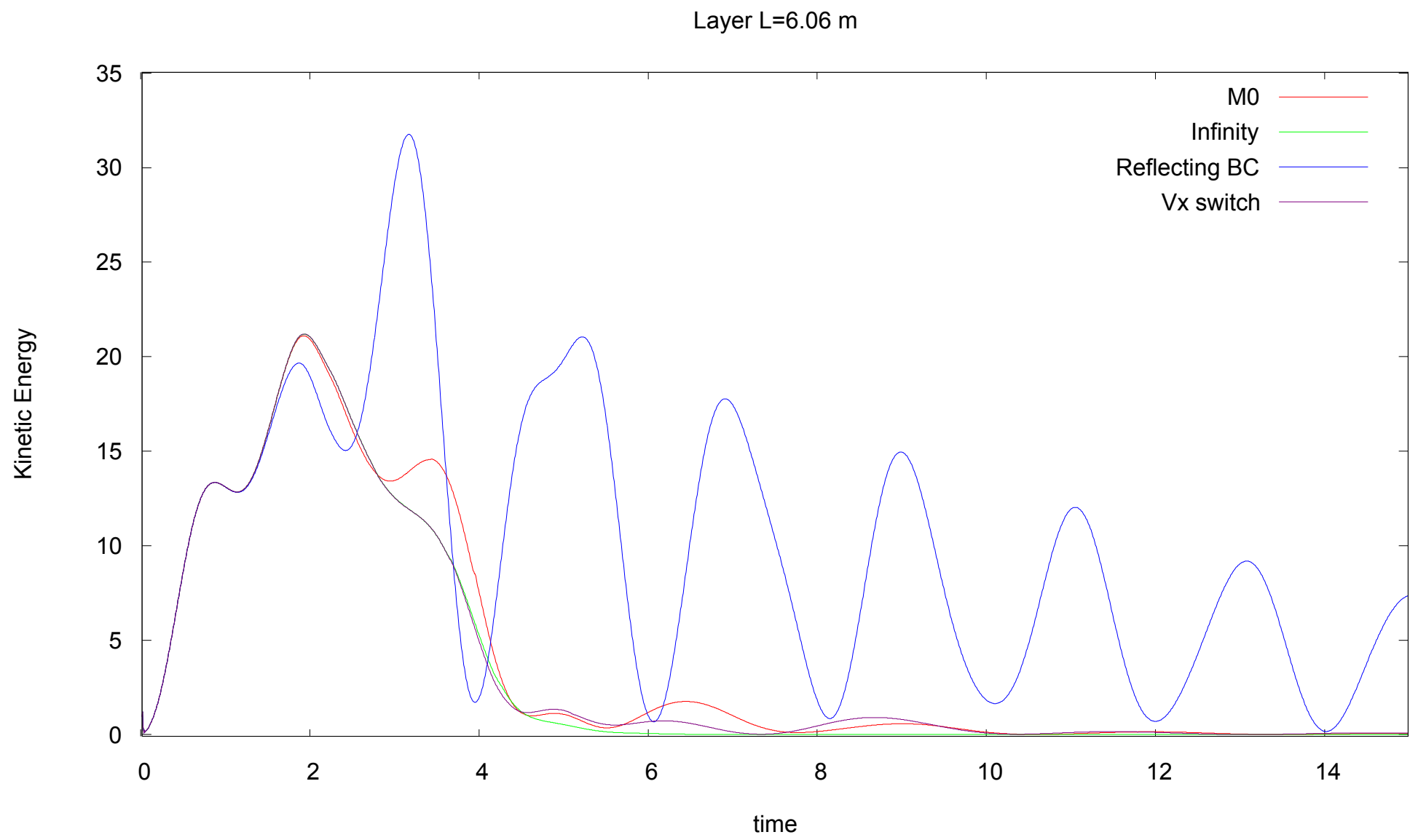

Fig. 18. Kinetic energy for different damping actions in the case of a 6.06 m large damping zone

Fig. 19 shows the kinetic energy of the water in the pool for a shorter damping layer L=3.03 m. We tested also the use of killing functions, but the improvements are very small to be appreciated in the figure. From this Fig. 19 it is clear the reducing action of the residual oscillations due to the velocity switch when added to the plain damping function.

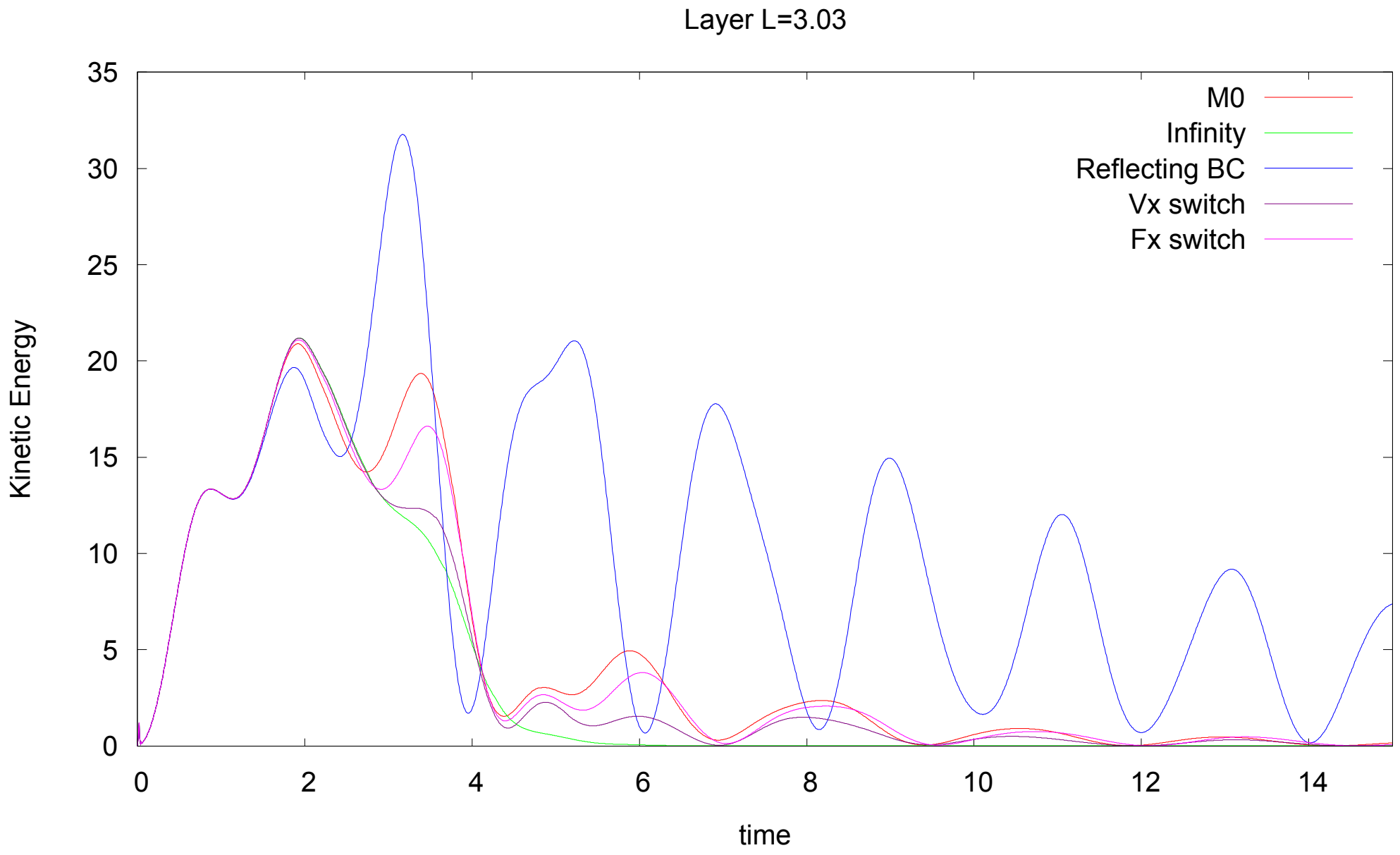

Fig. 19. Kinetic energy of the water in the case of damping layer L=3.03 m

Fig. 20 shows a zoom on the Fig. 19, to show clearly the different effects of the damping criteria. It shows that for the 2D problem the killing function improves the results over the plain damping, but not better than the simple velocity switch. The joint action of the velocity switch and of the killing function does not improve the result.

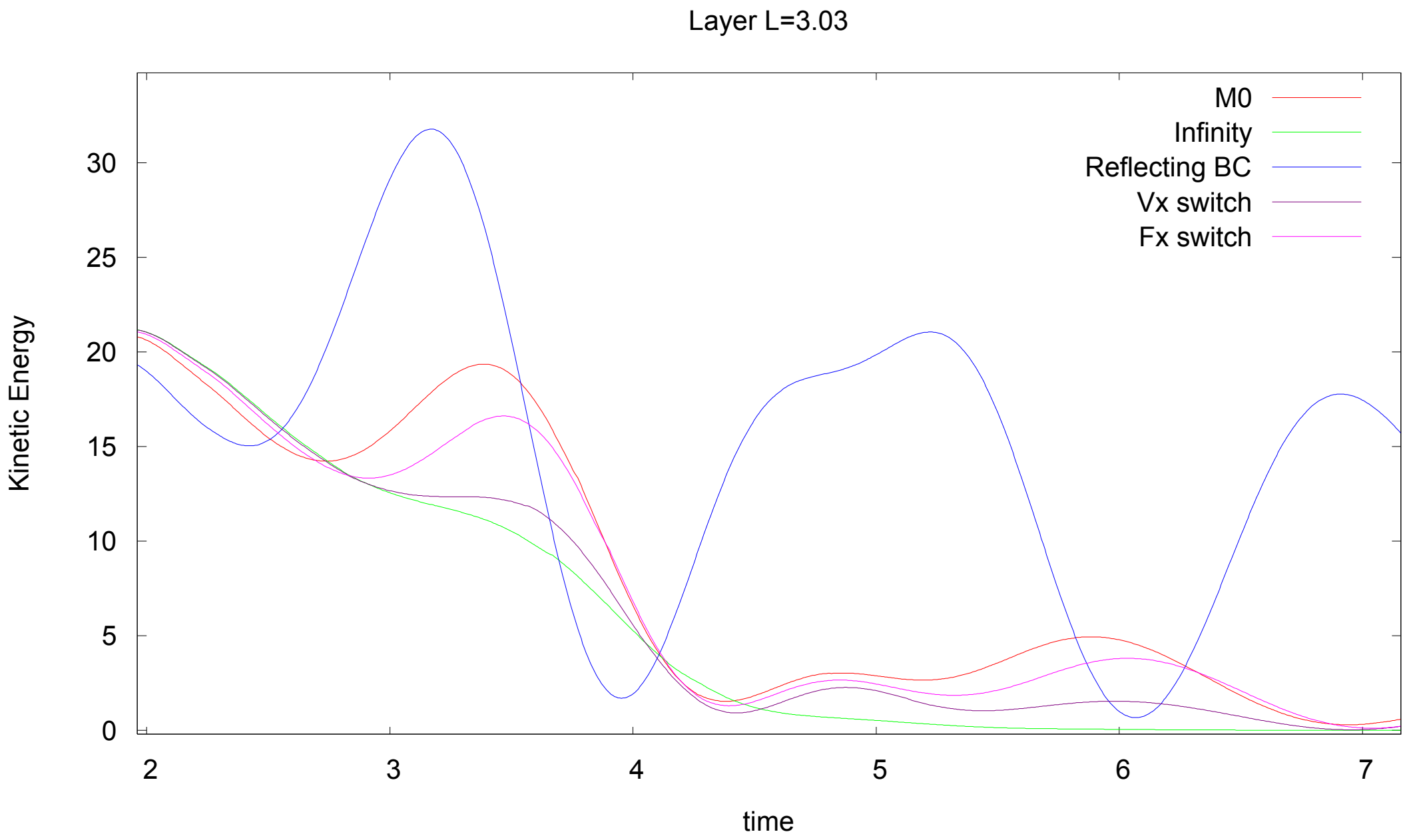

Fig. 20. Zoom on Fig. 19 to better show the results of the different damping algorithms

## 4 Conclusion

The use of using a damping layer is successful to avoid boundary reflections into the computational domain. An obvious requirement is that the absorbing layer must be greater or equal to the maximum significant wavelength produced by the physical simulation. Both in the 1D and 2D cases the basic procedure can be improved by the use of appropriate switches. A simple and efficient switch is the one that makes the damping to operate only for negative speeds, i. e. $v_x < 0$. The switch that reduces the horizontal component of the force, $F_x$, is efficient in 1D, but not so much for the 2D problem we studied. Further work is in progress to make a quantitative evaluation of the 2D simulations and verify the affordability of the method in the case of highly compressible fluid dynamics.

**Acknowledgements**

R. Grammauta acknowledges CRRSN for a six months grant exploited for this study.

**References**

Berenger, J.P., 1994. A Perfectly Matched Layer for the Absorption of Electromagnetic Waves. J. Comp. Phys. 114, 185-200.

Colagrossi, A., Landrini, M., 2003. Numerical simulation of interfacial flows by smoothed particle hydrodynamics. J. Comp. Phys. 191, 448–475.

Engquist, B., Majda, A., 1977. Absorbing boundary conditions for numerical simulation of waves, Proc. Natl. Acad. Sci. USA, Vol. 74, No. 5, pp. 1765-1766.

Lastiwka, M., Basa, M., Quinlan, N.J., 2009. Permeable and non-reflecting boundary conditions in SPH. Int. J. Numer. Meth. Fluids. 61, 709–724.

Lin, D.K., Li, X.D., Hu, Fang Q., 2011. Absorbing boundary condition for nonlinear Euler equations in primitive variables based on the Perfectly Matched Layer technique. Computer & Fluids. 40, 333-337.

Modave, A., Deleersnijder, É., Delhez, J.M., 2010. On the parameters of absorbing layers for shallow water models. Ocean Dynamic. 60,. 65–79.

Monaghan, J.J., 2005. Rep. Prog. Phys. 68, 1703-1759.

Monaghan, J.J., 2006. Mont. Not. R. Astron. Soc. 365, 199-213.

Poinsot, T. J., Lele, S.K., 1992. Boundary conditions for direct simulations of compressible viscous flows. J. Comp. Phys. 101, 104-29.